# Bounded Mean-Delay Throughput and Non-Starvation Conditions in Aloha Network

Soung Chang Liew, *Senior Member, IEEE*, Ying Jun (Angela) Zhang, *Member, IEEE*, Da Rui Chen, *Student Member, IEEE*

*The Department of Information Engineering, The Chinese University of Hong Kong*
*Shatin, New Territory, Hong Kong*
*Email: {scliew, yjzhang, drchen}@ie.cuhk.edu.hk*

***Abstract*—Prior investigations on the Aloha network has primarily focused on its system throughput. Good system throughput, however, does not automatically translate to good delay performance for the end users. Neither is fairness guaranteed: some users may starve while others hog the system. This paper establishes the conditions for bounded mean queuing delay and non-starved operation of the slotted Aloha network. We focus on the performance when collisions of packets are resolved using an exponential backoff protocol. For a non-saturated network, we find that bounded mean-delay and non-starved operation can be guaranteed only if the offered load is limited to below a quantity called "safe bounded-mean-delay (SBMD) throughput". The SBMD throughput can be much lower than the saturation system throughput if the backoff factor *r* in the exponential backoff algorithm is not properly set. For example, it is well known that the maximum throughput of the Aloha network is $e^{-1} = 0.3679$. However, for $r = 2$, a value assumed in many prior investigations, the SBMD throughput is only 0.2158, a drastic penalty of 41% relative to 0.3679. Fortunately, using $r = 1.3757$ allows us to obtain SBMD throughput of 0.3545, less than 4% away from 0.3679. A general conclusion is that the system parameters can significantly affect the delay and fairness performance of the Aloha network. This paper provides the analytical framework and expressions for tuning *r* and other system parameters to achieve good delay and non-starved operation.**

## 1. Introduction

The Aloha network has been studied extensively since the pioneer work by Abramson [1]. Prior work on the Aloha network has primarily focused on its overall system throughput. To achieve good system throughput, the transmission probabilities of the nodes must be adjusted dynamically according to the contention intensity in the network. An exponential backoff protocol can serve this purpose rather effectively [2].

Good system throughput, however, does not automatically translate to acceptable performance from the end user perspective. For example, if a real-time application such as a voice call is running on top the Aloha network, delay performance is important. Even if the end application is not a real-time application, there is also the fairness issue, wherein some nodes in the Aloha network are starved while other nodes enjoy good service. For example, for a TCP application, starvation could cause its TCP connection to terminate halfway.

This paper is devoted to the study of how to ensure good delay and non-starved performance in a slotted Aloha network operated with an exponential backoff protocol. In particular, we are interested in the setting of system parameters to attain not just good overall system throughput, but also good delay and fairness performance. Within this context, this paper has two major contributions:

1. We establish an analytical framework for the study of queuing delay and starvation in the Aloha network.

2. Based on the analytical framework, we derive the dependency of delay and non-starvation on the system parameters.

With respect to contribution 1, we unite the concepts of bounded mean-delay performance and non-starvation, arguing that the conditions giving rise to them are one of the same in a non-saturated Aloha network: namely, the service time at the heads of queues must be bounded. We find that the "saturation throughput", a performance metric of focus in many prior studies, is not a sound measure of performance if we care about delay and non-starved operation. In particular, to achieve good delay and non-starved operation, the offered load must be below another quantity called "safe-bounded-mean-delay (SBMD) throughput", which can be substantially lower than the saturation throughput. In establishing our analytical framework, we find that the delay analysis is much trickier than the saturation-throughput analysis in prior work. To better bring out the subtleties involved, we decompose our analysis into three steps: (i) a global analysis that captures the interaction among nodes; (ii) a local analysis that captures the dynamic within a node; (iii) a coupling analysis that integrates (i) and (ii) into a coherent whole. Steps (ii) and (iii) in the delay analysis, in particular, are a lot more involved than steps (ii) and (iii) in the saturation analysis.

With respect to contribution 2, we show that delay and non-starvation can be very sensitive to the system parameters; much more so than the saturation throughput is. For example, it is well known that the maximum saturation throughput of a large slotted Aloha network with many nodes is $e^{-1} = 0.3679$. An exponential backoff factor of 2 (see Section 2.1 for the definition of the backoff factor) was commonly assumed in many prior studies [3] [4]. It can achieve a saturation throughput of 0.3466 [2]. Thus, backoff factor of 2 is quite satisfactory when it comes to saturation throughput performance. However, if we desire bounded mean-delay and non-starved performance, we must limit the system offered load to below 0.2158, a drastic penalty of 41% with respect to the maximum throughput. Fortunately, if instead of 2, we use a backoff factor of 1.3757, the sustainable offered load can reach 0.3545, very close to the maximum throughput. This paper will present many other intricate relationships between system parameters and system operation and performance.

**Related Work**

Most prior investigations on the Aloha network (e.g., [2][3][5]) consider the access delay (i.e., service time incurred by a packet at the head-of-line (HOL) of its queue). Less attention is paid to the overall queuing delay (i.e., waiting time plus service time). In [2] the saturation throughput (reciprocal of mean access delay) as a function of the backoff factor *r* was derived. A fundamental expression obtained in [2] is the dependency of saturation throughput on *r* for a large network: $S_s = \frac{r-1}{r}\ln\left(\frac{r}{r-1}\right)$. Higher moments of the access delay, however, were not considered. Ref. [3] focused on the case of $r=2$ only and investigated both the mean and variance of access delay. It was shown that the throughput must be below $3(\ln 4 - \ln 3)/4 = 0.2158$ if variance of access delay is to be finite.

In contrast to these prior investigations, a focus of our work here is on the queuing delay rather than the access delay, and for general *r*. For the $r=2$ case, bounded mean delay requires only the access-delay variance to be bounded. Hence the sustainable offered load for bounded mean queuing delay is the same as that derived in [3] for bounded access-delay variance. For *r* smaller than 1.3757, however, we argue in this paper that an offered load that ensures bounded access-delay variance cannot safely guarantee bounded mean queuing delay, and that the offered load must also be below the saturation throughput.

As in this paper, [6] also considered the non-saturated scenario, but for 802.11 networks. Furthermore, the focus is on throughput rather than on the delay performance. It argued that the notion of saturation throughput is a pessimistic one in that the system throughput could be above the saturation throughput if the queues are forced to be emptied from time to time. We find that as far as the Aloha network is concerned, with an appropriate setting of *r*, one could achieve throughput that is only less than 4% away from the maximum throughput of $e^{-1}$. This is achieved without forced emptying of queues, and with delay performance taken into consideration.

In this paper, we consider a slightly different exponential backoff protocol than the prior work. Our model captures the main essence and principle of exponential backoff and has the advantage of being more amenable to analysis. Many of the saturation throughput results in [2] can be obtained within the space of less than one page with our model, as will be shown in Section 2.1.

We are primarily interested in networks in which the number of nodes *N* is large. Our large-but-fixed-*N* results are not to be confused with the results of the infinite-population model [7] in which nodes, each with one and only one packet to transmit, is created on the fly. In the former, the number of contending packets is bounded by *N*, whereas in the latter, the number of contending packets can grow indefinitely. As a matter of fact, the saturation throughput of binary exponential backoff is 0.3466 in the limit of $N \to \infty$ in the former, but zero in the latter [8].

The remainder of this paper is organized as follows. Section 2 presents our system model. We illustrate the use of the model in saturation analysis. Many expressions useful for queuing-delay and starvation analyses later are derived. Section 3 presents our queuing-delay analysis. We derive expressions that relate delay performance to system design parameters. The materials presented in Section 3 show that queuing-delay analysis is much more subtle than the saturation-throughput analysis in Section 2 and in prior work. Section 4 investigates in detail the effects of the backoff factor on the sustainable offered load for bounded mean-delay operation. Section 5 is devoted to the study of the starvation phenomenon. We derive the dependency of starvation on system parameters. Section 6 concludes this paper.

## 2. System Model and Saturation Analysis

In this section, we first describe the system model under study, and then perform a saturation analysis.

### 2.1 System Model

**Real System**

We consider a slotted Aloha network with *N* nodes. Each node has a queue to hold its backlog packets. When a fresh packet enters the HOL of its queue, it transmits with probability $1/r_0$ in each time slot, where $r_0 \ge 1$. When more than one node transmits a packet in a time slot, a collision occurs and the packets are corrupted. A collided packet will be retransmitted in a future time slot. Each time a HOL packet suffers a collision, the transmission probability in the future is divided by the backoff factor $r > 1$. Thus, a HOL packet that has suffered *i* prior collisions will be transmitted in a future time slot with probability $1/(r_0 r^i)$. We refer to *i* as the backoff stage of a node. A HOL packet will be transmitted and retransmitted until it is successfully cleared without a collision, at which point the next-in-line packet, if any, will proceed to the HOL.

Another closely related protocol often considered is that of a countdown-window protocol [2][9] in which a countdown process is used to determine when a HOL packet is transmitted. The parameter $r_0$ in our protocol serves the same purpose as the initial window size $W_0$ of that model in determining the expected number of time slots until the first transmission of a HOL packet; and the common backoff factor *r* serves the same purpose in both models: i.e., for dynamic adjustment of the transmission probabilities of nodes according to contention intensity. For a given *r*, the two protocols have roughly the same behavior if $r_0 \approx W_0/2$. Our model, however, is simpler to analyze. In Section 2.1, for example, we show that many saturation results similar to those in [2] can be obtained in a few simple steps within the space of less than a page.

With our model, the "local state" of a queue can be described by a duple $(Q,B)$, where *Q* is the number of backlog packets in the queue, including the HOL packet; and $B$ is the backoff stage of the HOL packet. The "global state" of the overall system consists of the aggregate local states of all $N$ queues. One can in principle construct a multi-dimensional Markov chain to analysis the system. However, the analysis for even modest-size $N$ is prohibitively complex and not much insight can be gained from this brute-force analysis. Detailed and exact results, for example, are only available for the 2-node case [4].

**Proxy System**

For large $N$, an approximation technique that has been often used in saturation analysis is to replace the actual system model with a "proxy model" (e.g., used in [2], as well as [9] and many of its follow-up papers). This paper adopts the same approximations for saturation as well as non-saturation analyses.

The proxy system makes two approximations: (i) the probability of collision $p_c$ experienced by a node is independent of its local state; (ii) as far as a local node is concerned, each of the other nodes transmits with a probability $p_t$ in a given time slot. Certainly these approximations are only valid under large $N$ when each local node only has a small effect on the overall system. Simulations of the actual system, referred to as the "real system" in this paper, can be used to check against the accuracy of the proxy-system analysis. This paper will show such verification results.

In this paper, for better exposition and understanding of the intricacies involved, we decompose the analysis of the proxy system into three steps. The first step is a "global analysis" linking $p_c$ and $p_t$: *viz* $p_c = 1-(1-p_t)^{N-1}$. The second step is a "local analysis" focusing on the local dynamic of a node assuming a fixed $p_c$. The third step is a "coupling analysis" which combines the results from the first two steps to obtain $p_c$ in terms of system parameters $r_0, r, N$.

## 2.2 Saturation Throughput Analysis

We now illustrate the three-step analytical technique for the proxy system by performing a saturation analysis. Besides illustrating the three-step technique, more importantly, the reason for going through the motion to establish some of the saturation results here is that they will be used later as part of our delay analysis (Sections 3 and 4) and starvation analysis (Section 5).

**Global Analysis**

Consider the overall system consisting of the $N$ homogenous nodes. Recall that $p_t$ is the probability of transmission of an arbitrarily chosen node in the proxy system, and $p_c$ is the collision probability of a transmitting node. By the homogeneity assumption of the proxy system, we have

$$\begin{aligned} p_c &= 1-(1-p_t)^{N-1} \\ &\to 1-e^{-Np_t} \quad \text{as } N\to\infty \end{aligned} \tag{1}$$

Define $G_s = Np_t$ as the global transmission attempt rate, and $S_s$ as the saturation throughput of the overall system. Then, by definition,

$$\begin{aligned} S_s &= G_s(1-p_c) = G_s\left(1-\frac{G_s}{N}\right)^{N-1} \\ &\to G_s e^{-G_s} \quad \text{as } N\to\infty \end{aligned} \tag{2}$$

The expression $S_s = G_s e^{-G_s}$ for the asymptotic case, of course, is the well-known slotted Aloha throughput equation. The relationships in (2) govern the global dynamic of the system.

**Local Analysis**

Consider one particular node. Let $X$ be the HOL access delay of a packet. Then, by considering the successive additional expected access delays incurred conditioned on the number of collisions, we have $E[X] = r_0 + r_0 r p_c + r_0 r^2 p_c^2 + \ldots = r_0/(1-rp_c)$. At saturation, the HOL is always occupied. Hence, by Little's Law, we have $E[X]S_s/N = 1$, where $S_s/N$ is the saturation throughput of the local node. These two equations give

$$\begin{aligned} p_c &= \frac{1}{r}\left(1-\frac{r_0 S_s}{N}\right) \\ &\to \frac{1}{r} \quad \text{as } N\to\infty \end{aligned} \tag{3}$$

**Coupling Analysis**

We now couple the results from the global and local analyses. Overall, we can express any of the variables $S_s$, $G_s$, $p_c$, or $p_t$ in terms of the system parameters $r, r_0, N$. In the following, we only list the expressions that will be used later.

The dependency of $p_c$ on system parameters $r, r_0, N$ will be useful for our starvation analysis later. From $S_s = G_s(1-p_c)$, $S_s = G_s\left(1-G_s/N\right)^{N-1}$ in (2) and $p_c = (1-r_0 S_s/N)/r$ in (3), we can get

$$1-p_c = \left(1-\frac{(1-p_c r)}{r_0(1-p_c)}\right)^{N-1} \tag{4}$$

The same three equations also give us

$$\begin{aligned} r-1 &= \left(1-\frac{G_s}{N}\right)^{N-1}\left(r-\frac{r_0 G_s}{N}\right) \\ \left(1+\frac{r_0}{r-1}\cdot\frac{S_s}{N}\right)^{N} &= \left(\frac{r}{r-1}\right)\left(1+\frac{r_0-r}{r-1}\cdot\frac{S_s}{N}\right)^{N-1} \end{aligned} \tag{5}$$

For $N\to\infty$, $S_s = G_s(1-p_c)$, $S_s = G_s e^{-G_s}$ in (2) and $p_c = 1/r$ in (3) yield (below can also be obtained by taking limit in (5))

$$\begin{aligned} G_s &= \ln\left(\frac{r}{r-1}\right) \\ S_s &= \left(\frac{r-1}{r}\right)\ln\left(\frac{r}{r-1}\right) \end{aligned} \tag{6}$$

Note that while the solution for $S_s$ is in closed form in the asymptotic case, $S_s$ must be found numerically from (5) in the finite-$N$ case. Also, $S_s$ depends on $r_0$, $r$, $N$ in the finite-$N$ case but only on $r$ in the asymptotic case. The practical significance of (5) and (6) is that they allow us to study the dependency of the saturation throughput $S_s$ on system parameters $r, r_0, N$.

## 3. Delay Analysis

We now consider the non-saturation analysis in which the queues of the nodes are not saturated. Unless otherwise stated, henceforth by "delay" we mean "queuing delay" rather than the "HOL access delay". We assume the arrival process to each queue is Poisson with rate $\lambda_o = S_o/N$, where $S_o$ is the offered load to the overall system, and thus $S_o/N$ is the offered load to a single queue.

For a non-saturated system under equilibrium, the output rate (i.e., throughput) is equal to the input rate (i.e., offered load). Given a system with system parameters $r, r_0, N$, we could load it with different offered load $S_o$, and therefore obtain different throughput $S_o$. This is in contrast to a saturated system, in which the saturation throughput $S_s$ is a "fixed" quantity given $r, r_0, N$.

Different $S_o$, however, will give rise to different delay performance, and it is important not to overload the system. An issue of particular interest to us, which will be addressed by the end of this section, is the limit on $S_o$ that can ensure equilibrium and bounded-delay operation. We call this limit "safe-bounded-mean-delay throughput". As will be shown, safe-bounded-mean-delay throughput depends on $r, r_0, N$ and may be lower than $S_s$.

As with the saturation analysis in Section 2.2, we break down the delay analysis into global, local, and coupling analyses. It turns out that the local and coupling analyses are much more involved here.

### 3.1 Global Analysis

The global analysis of throughput is largely the same as that of the saturated system given the two approximations of the proxy system described in Section 2.1. That is, (1) and (2) remain valid with the replacements of $S_s$ by $S_o$ and $G_s$ by $G_o$, where $G_o = Np_t$ is the transmission attempt rate of the overall system when the offered load (throughput) is $S_o$. Parallel to (2), we have

$$\begin{aligned} S_o &= G_o\left(1-\frac{G_o}{N}\right)^{N-1} \\ &= G_o e^{-G_o} \qquad \text{as } N \to \infty \end{aligned} \tag{7}$$

where $G_o = Np_t$ is the transmission attempt rate of the overall system when the offered load is $S_o$.

### 3.2 Local Analysis

The local analysis is more complicated than that in the saturated case, since we need to consider the queuing dynamic at a node, not just the HOL contention dynamic. For Poisson arrival, a packet of a local queue generally arrives between the boundaries of two adjacent time slots. If it arrives to an empty queue, it must wait until the beginning of the next time slot before it can contend for transmission. Conceptually, it does not enter the HOL until the next time slot. It turns out that this local queue specification fits under the M/G/1 multiple-vacation queue model [10], as elaborated in the next paragraph. The intricate part of our analysis is in deriving the service-time distribution and the vacation-time distribution of the Aloha system to substitute into the equations of the M/G/1 vacation queue.

In the multiple-vacation queue model [10], the server may leave for a vacation when the queue becomes empty. The vacation length is a random variable *V*. Upon returning from a vacation, if the queue remains empty, the server immediately departs for another vacation. When a packet arrives to an empty queue in the Aloha network, the time until the beginning of the next time slot is part of the vacation time taken by the server. For slotted Aloha, the vacation time is fixed and equal to one slot time. The access delay incurred by a packet at the HOL corresponds to the service time of the M/G/1 vacation queue model.

For notation purposes, in the following, $F(z) = \sum_{i=0}^{\infty} \Pr[F = i] z^i$ denotes the *z*-transform of a discrete non-negative random variable *F*, and $G^*(s) = \int_0^{\infty} f_G(x) e^{-sx} dx$ denotes the Laplace transform of a continuous non-negative random variable *G*. The M/G/1 vacation queue has the following solution:

$$Q(z) = \frac{(1-\lambda_o \bar{X})}{\lambda \bar{V}} \cdot \frac{X^*(\lambda_o(1-z))[V^*(\lambda_o(1-z))-1]}{z - X^*(\lambda_o(1-z))}$$

$$D^*(s) = Q(1-s/\lambda_o) = \frac{(1-\lambda_o \bar{X})}{\bar{V}} \cdot \frac{X^*(s)[V^*(s)-1]}{\lambda - s - \lambda_o X^*(s)}$$

where

$Q$ = number of packets in the queue including the HOL packet (8)

$D$ = queueing delay including the service time

$X$ = service time of a packet

$V$ = vacation time taken by the server when the queue is empty

Expressions (8) are generic expressions relating $Q$ and $D$ to $X$ and $V$. To use (8), however, we need to derive the distributions of $X$ and $V$ specific to our system. For slotted Aloha, each vacation lasts exactly one time slot, so that

$$V^*(s) = e^{-s} \tag{9}$$

Recall that an approximation in the proxy system is a constant $p_c$ independent of the local state. We now derive $X$ in terms of $p_c$. Mathematically, the Laplace transform $X^*(s)$ in (9) is related to the *z*-transform $X(z)$ by

$$X^*(s) = X(e^{-s}) \tag{10}$$

To derive $X(z)$, let $C$ be the number of collisions experienced by a HOL packet before it is successfully transmitted. By conditional-probability argument, we have

$$X(z) = \sum_{k=0}^{\infty} X(z \mid C=k)(1-p_c)p_c{}^k,$$

$$X(z \mid C=k) = X_0(z)X_1(z)\cdots X_k(z)$$

where

$$X_j(z) = z\text{-transform of the time between the } j^{\text{th}} \text{ and } (j+1)^{\text{th}} \text{ transmissions} \tag{11}$$

$$= \sum_{i=1}^{\infty}\left(\frac{1}{r_0 r^j}\right)\left(1-\frac{1}{r_0 r^j}\right)^{i-1} z^i = \frac{z}{r_0 r^j - (r_0 r^j - 1)z}$$

Thus,

$$X(z) = \sum_{k=0}^{\infty}(1-p_c)p_c{}^k \prod_{j=0}^{k}\frac{z}{r_0 r^j - (r_0 r^j - 1)z} \tag{12}$$

Eqns. (8), (9), (10), and (12) allow us to derive moments of $D$ in terms of $r_0$, $r$, $p_c$. For the first moment $E[D]$, after some equation crunching, we can get

$$\begin{aligned} E[D] = -D^{*}{}'(0) &= X'(1) + \frac{\lambda_o X''(1)}{2(1-\lambda_o X'(1))} + \frac{\lambda_o X'(1)}{2(1-\lambda_o X'(1))} + \frac{\overline{V^2}}{2\overline{V}} \\ &= \frac{r_0}{1-p_c r} + \frac{\lambda_o r_0 (p_c r^2 + r_0 - 1)}{(1-p_c r^2)(1-p_c r - \lambda_o r_0)} + \frac{\lambda_o r_0}{2(1-p_c r - \lambda_o r_0)} + \frac{1}{2} \\ &= \frac{r_0}{1-p_c r} + \frac{\lambda_o r_0 (p_c r^2 + 2r_0 - 1)}{2(1-p_c r^2)(1-p_c r - \lambda_o r_0)} + \frac{1}{2} \end{aligned} \tag{13}$$

We note that independently [11] obtained $X'(1)$ and $X''(1)$ for the $r_0 = 1$ case. Let us next consider the implications of (13).

**Bounded Mean-Delay Conditions**

We focus on the conditions to ensure bounded mean delay in the following. As mentioned above, higher moments of delay, such as delay variance can also be obtained from (8), (9), (10), and (12) in principle. If desired, argument similar to that below can also yield the conditions for bounded delay variance.

From (13), convergence of $E[D]$ requires $p_c r < 1$, $p_c r + \lambda_o r_0 < 1$ and $p_c r^2 < 1$, but the first inequality is satisfied if the second is and can be eliminated. Thus, we have the following conditions for convergence:

$$p_c r + \lambda_o r_0 = p_c r + \frac{r_0 S_o}{N} < 1 \quad \text{and} \quad p_c r^2 < 1 \tag{14}$$

Note that at equilibrium, the mean service time is $X'(1) = r_0/(1-p_c r)$. Applying Little's law and requiring the average HOL occupancy to be less than 1, we have $\lambda_o r_0/(1-p_c r) < 1$, which is the same as the first inequality in (14). Thus, the first inequality is also the condition for non-saturation.

The analysis thus far assumes steady-state equilibrium can be achieved. For a queuing system, steady state can be achieved if and only if $\Pr[Q=0] > 0$ (see [12]). Since $\Pr[Q=0] > 0$ means the queue is not saturated, the first inequality of (14) is also the necessary and sufficient condition for steady state operation. In other words, non-saturated operation is the same as steady-state operation.

The second inequality in (14) arises from the requirement to bound $Var(X) = X''(1) + X'(1)$ in (13). We note that unbounded $E[D]$ does not automatically imply that the system is saturated, although the converse is true. To see this, consider a *hypothetical* distribution of $Q$ that does not decay fast enough: $\Pr[Q=0] = 1/2$, $\Pr[Q=i] = 3/(\pi i)^2$ for $i \ge 1$. It is easy to see that $E[Q] = \sum_i i \Pr[Q=i]$ (hence $E[D]$ also) is unbounded, but the system is not saturated because $\Pr[Q=0] \ne 0$.

In short, bounded $E[D]$ requires both the system to be non-saturated (first inequality in (14)) and the variance of the service time to be bounded (second inequality in (14)).

## 3.3 Coupling Analysis

The coupling analysis also involves many subtleties not present in the saturation case. The local analysis leaves us with (13), where mean delay is expressed in terms of $r$, $r_0$ and $p_c$. We need to use the result from the global analysis to remove the dependency on $p_c$. As elaborated below, in doing so, we find ourselves in the quandary of having two possible $p_c$, which in turn give rise to two possible $E[D]$. We explore this subtlety below and argue that only one of the two possible $p_c$ is valid upon closer examination.

### 3.3.1 Quantum Jump of Equilibrium Operating Point

For exposition purposes, we consider the asymptotic $N \to \infty$ case here. Similar argument applies to the finite-$N$ case. First, we note that for saturated operation, for each fixed $S_s$, there are two possible $G_s$ according to the global-analytical result $S_s = G_s e^{-G_s}$ from (2). These two $G_s$ correspond to two different backoff factors $r$ according to the coupling-analytical result (6). That is, two different $r$ can be used to achieve the same $S_s$ and they have different $G_s$. Fig. 1 is a pictorial illustration. The two $G_s$ are $G_l$ on the left and $G_r$ on the right, and the corresponding two $r$ are $r_l$ and $r_r$, respectively. From (6) we know that $G_s$ is an decreasing function of $r$, and therefore $r_l \ge r_r$.

Right after (14), we argued that the system must not be saturated in order that equilibrium can be achieved. Suppose that we load the system with $S_o < S_s$ to ensure non-saturated operation. Consider the two systems with $r_l$ and $r_r$, respectively. The global $S_o$-versus-$G_o$ and $S_s$-versus-$G_s$ curves have the same form: $S = Ge^{-G}$. So, we can overlay the saturation and non-saturation operating points on the same graph, as in Fig. 1. As shown in Fig. 1, for the given $S_o < S_s$, we could draw a horizontal line below $S_s$ to identify the corresponding $G_o$. We find that for the given $S_o$, we have two possible $G_o$: $G_{o,l}$ and $G_{o,r}$ with $G_{o,l} < G_{o,r}$. Which of them is the "correct" operating point?

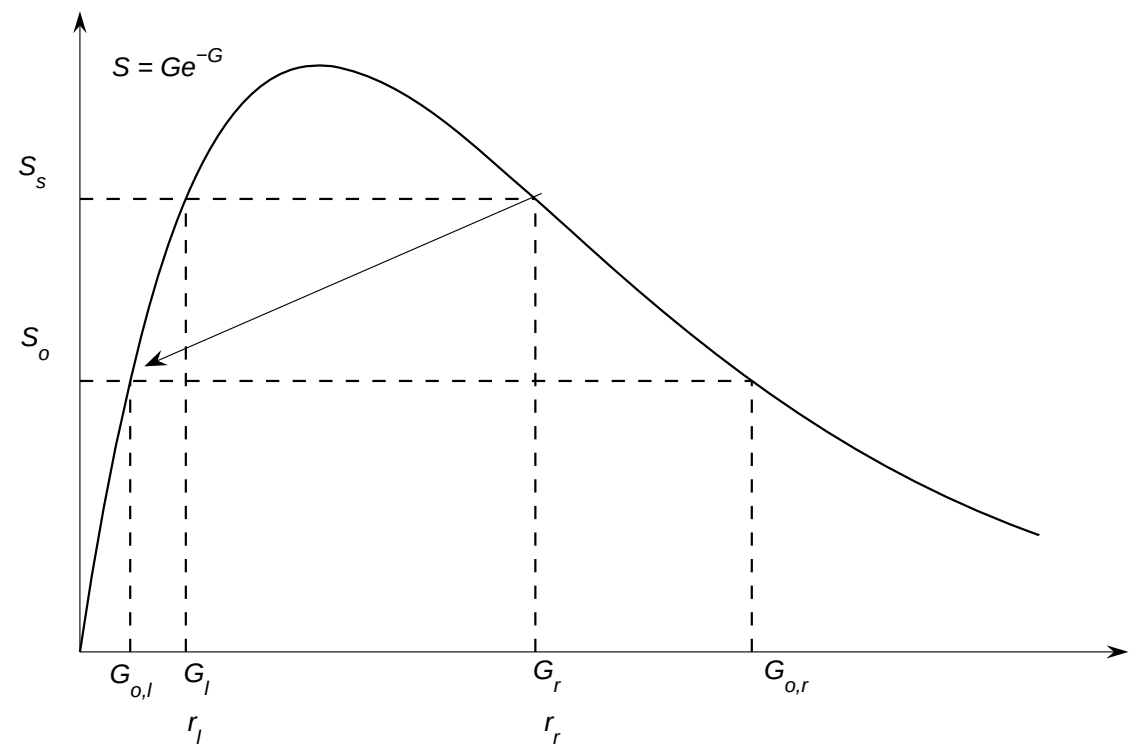

Fig. 1. Illustration of the quantum jump in operating point when $r_r$ is used.

It is tempting to jump into the conclusion that in the system with $r_l$ , $G_{o,l}$ is the operating point; and in the system with $r_r$ , $G_{o,r}$ is the operating point. After all, this gives a smooth and continuous transition from the two operating points at saturation, $G_l$ and $G_r$, as $S_o$ is decreased slowly from $S_s$. It turns out that this is not the case. As argued below, when the system is not saturated, the operating point is $G_{o,l}$ for both $r_l$ and $r_r$; the operating point $G_{o,r}$ is not tenable for either $r_l$ or $r_r$

For the systems with $r_l$ and $r_r$ , by definition their $p_c$ at saturation are $p_c = 1 - S_s/G_l$ and $p_c = 1 - S_s/G_r$ , respectively. From the second line of (3), $p_c r = 1$ at saturation; thus, we have $(1 - S_s/G_l)r_l = 1$ and $(1 - S_s/G_r)r_r = 1$ for $r_l$ and $r_r$, respectively.

At offered load $S_o$, the $p_c$ at the "potential" operating points $(G_{o,l}, S_o)$ and $(G_{o,r}, S_o)$ are $p_c = 1 - S_o/G_{o,l}$ and $p_c = 1 - S_o/G_{o,r}$ , respectively. If $(G_{o,r}, S_o)$ were the operating point under $r_l$ and $r_r$ , we would have respectively the following:

$$\begin{aligned} p_c r_l &= \left(1 - S_o/G_{o,r}\right) r_l > \left(1 - S_s/G_l\right) r_l = 1 \\ p_c r_r &= \left(1 - S_o/G_{o,r}\right) r_r > \left(1 - S_s/G_r\right) r_r = 1 \end{aligned} \tag{15}$$

The inequalities in (15) can be seen as follows. Since $G_{o,r} > G_l$ and $G_r$ , we have $S_o/G_{o,r} = e^{-G_{o,r}} < e^{-G_l} = S_s/G_l$ ; and $S_o/G_{o,r} = e^{-G_{o,r}} < e^{-G_r} = S_s/G_r$ . Inequalities (15) imply that $(G_{o,r}, S_o)$ cannot be the operating point under $r_l$ or $r_r$ because $p_c r_l > 1$ and $p_c r_r > 1$ violate the condition for non-saturated and equilibrium operation (see (13) and argument leading to (14) and thereafter).

By contrast, the operating point at $(G_{o,l}, S_o)$ satisfies $p_c r < 1$ for both $r_l$ and $r_r$, as can be seen from below:

$$\begin{aligned} p_c r_l &= \left(1 - S_o/G_{o,l}\right) r_l < \left(1 - S_s/G_l\right) r_l = 1 \\ p_c r_r &= \left(1 - S_o/G_{o,l}\right) r_r < \left(1 - S_s/G_r\right) r_r = 1 \end{aligned} \tag{16}$$

We therefore conclude that the correct $G_o$ is the smaller of the two possible solutions to $S_o = G_o e^{-G_o}$. Note in particular that it does not matter what *r* is. The value of *r* only determines the saturation throughput $S_s(r)$. As long as we load the system with an offered load $S_o$ smaller than $S_s(r)$, $G_o$ is independent of $r$.

In Fig. 1, note also that for $r_r$, as we decrease $S_o$ from $S_o = S_s$ to $S_o < S_s$ (i.e., moving from saturation operation to non-saturation operation), there is a quantum jump in the transmission attempt rate from $G_s$ to $G_o$ (hence from $p_c = 1 - S_s/G_s$ to $p_c = 1 - S_o/G_o$ ), as illustrated by the arrow in Fig. 1. We have performed simulations on the "real system" to verify this analytical conclusion. Fig. 2 shows the simulation results in which $(N, r_0) = (20,10)$ , and $r = 1.04,\ 1.06,\ ...,\ 1.2$, which corresponds to $r_r$. The right curve is the saturation throughput $S_s$ versus $G_s$ curve. The left curve is the $S_o$ versus $G_o$ curve when we load the network with $S_o = 0.9S$ for each of the *r*. The quantum jumps predicted analytically by the proxy system are obvious from the simulation results of the real system. We summarize our finding in Observation 1 below.

**Observation 1:** For a given set of system parameters $r,\ r_0,\ N$ , if the resulting $(G_s, S_s)$ lies to the left of the peak of the *S*-*G* curve in (2), then the feasible non-saturated operating region is all points $(G_o, S_o)$ to the left of $(G_s, S_s)$ on the *S*-*G* curve. On the other hand, if $(G_s, S_s)$ lies to the right of the peak of the *S*-*G* curve in (2), then the feasible non-saturated operating region is all points to the left of $(G_l, S_s)$, where $(G_l, S_s)$ is the point to the left of the peak with the same saturation throughput $S_s$.

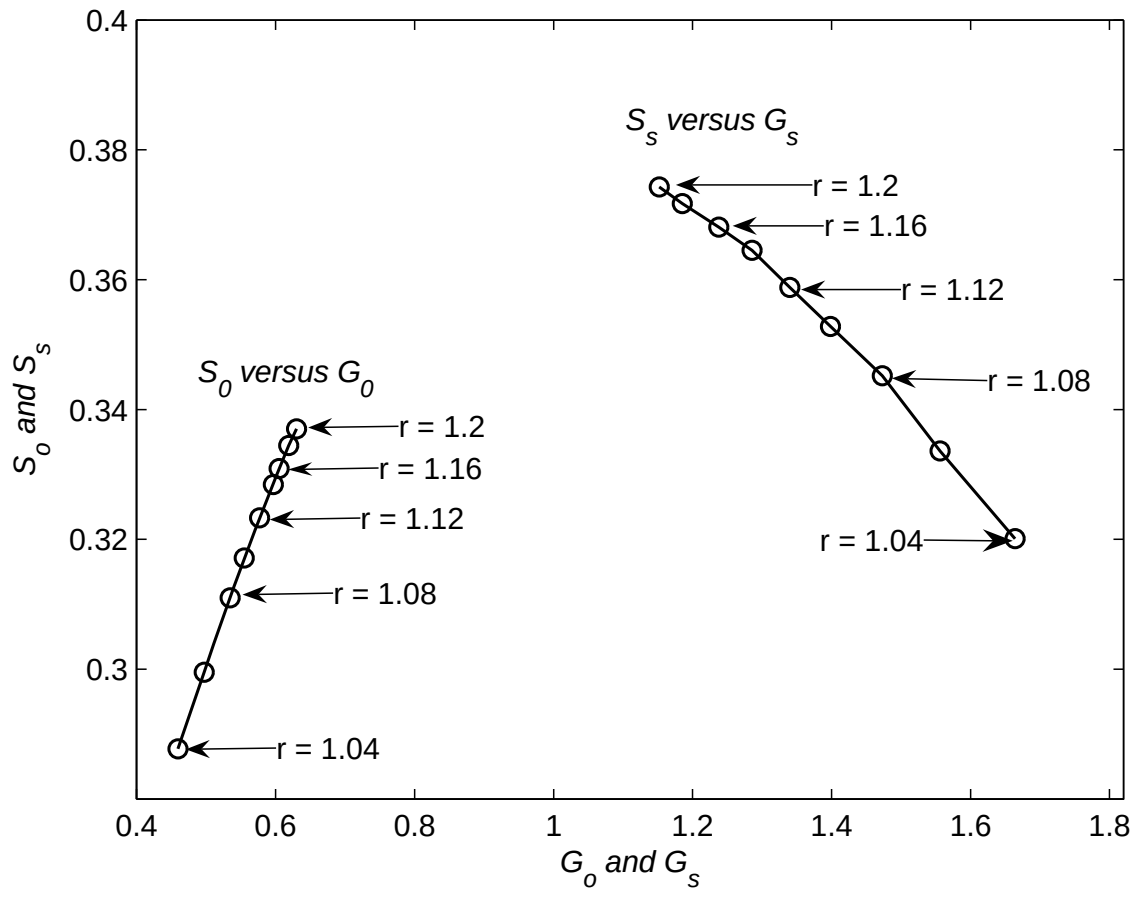

Fig. 2. Simulation results verifying the quantum-jump phenomenon of the saturation and below-saturation operating points.

### 3.3.2 Safe-Bounded-Mean-Delay (SBMD) Throughput $S_{SBMD}$

It turns out that there are further subtleties in the coupling analysis. Although having $S_o < S_s$ will ensure non-saturated operation, depending on the system parameters, this may or may not be sufficient for ensuring bounded-delay operation. That is, the feasible region established in Observation 1 pertains to non-saturated operation only. To bound mean delay, we explain in the following that additionally $S_o$ cannot exceed another value, $S_{BBMD}$, which we refer to as the bounded mean-delay throughput.

Specifically, $S_o$ must be smaller than the minimum of $S_s$ and $S_{BBMD}$. We refer to $S_{SBMD} = \min[S_{BBMD}, S_s]$ as the safe-bounded-mean-delay throughput.

The practical significance is as follows. If we load the system with $S_o > S_{SBMD}$, the system will have unbounded mean delay. When that happens, one or both of the following may occur: (i) queues may become saturated and/or delay may become unbounded; (ii) different queues may experience widely different performance even though they all operate the same protocol and have the same homogeneous offered load $S_o/N$. Issue (ii) will be discussed in more detail under the context of "starvation" in Section 5. We first expound on the concept of $S_{SBMD}$ here.

**$N \to \infty$ Case : $S_{SBMD}$ as function of *r***

For simplicity, we first consider the asymptotic $N \to \infty$ case. For $N \to \infty$, the first inequality in (14) becomes $p_c r < 1$. In particular, the first inequality is satisfied if the second inequality $p_c r^2 < 1$ is satisfied, since $r > 1$. Thus, we only need to look at the second inequality of (14).

Suppose we look at the "boundary" where $p_c r^2 = 1$. This is the boundary operating point where the mean delay goes to infinity. Note that $p_c$ varies on the $S_o = G_o e^{-G_o}$ curve according to $p_c = 1 - e^{-G_o}$. Using this fact on $p_c r^2 = 1$, we can get $S_o = \frac{r^2 - 1}{r^2} \ln\left(\frac{r^2}{r^2 - 1}\right)$. We shall refer to this quantity as the boundary-bounded-mean-delay throughput, denoted by

$$S_{BBMD} = \frac{r^2 - 1}{r^2} \ln\left(\frac{r^2}{r^2 - 1}\right) \tag{17}$$

The corresponding attempt rate is

$$G_{BBMD} = \ln\left(\frac{r^2}{r^2 - 1}\right) \tag{18}$$

Recall that (14) is obtained from local analysis, and therefore (17) an (18) are outcomes of local analysis. The local analytical results (17) and (18) dictate which of the operating points on the global-analytical curve $S_o = G_o e^{-G_o}$ are feasible and which are not for bounded mean-delay operation.

**Observation 2:** For a given *r*, bounded mean delay requires the operating point $(G_o, S_o)$ to lie to the left of $(G_{BBMD}, S_{BBMD})$ on the $S_o = G_o e^{-G_o}$ curve.

To see the validity of Observation 2, note that $p_c = 1 - e^{-G_o}$ and therefore $p_c$ increases with $G_o$. Thus, for a given *r*, in order that an operating point $(G_o, S_o)$ has $p_c r^2 < 1$, it must lie to the left of $(G_{BBMD}, S_{BBMD})$, where $p_c r^2 = 1$.

**Observation 3:** For a given *r*, $(G_s, S_s)$ is always to the right of $(G_{BBMD}, S_{BBMD})$ on the $S = Ge^{-G}$ curve.

To see the validity of Observation 3, note from (6) and (18) that $G_s = \ln\frac{r}{r-1} > \ln\frac{r^2}{(r-1)(r+1)} = G_{BBMD}$.

To identify the feasible region for bounded mean delay and non-saturated operation, in Fig. 3(a) to Fig. 3(d), we trace the movement of $(G_{BBMD}, S_{BBMD})$ according to (17) and (18), and the movement of $(G_s, S_s)$ according to (6), as *r* decreases. Both points move to the right as *r* decreases. The darkened lines in Fig. 3(a) to Fig. 3(d) correspond to the feasible operating regions. We explain each of the four cases below.

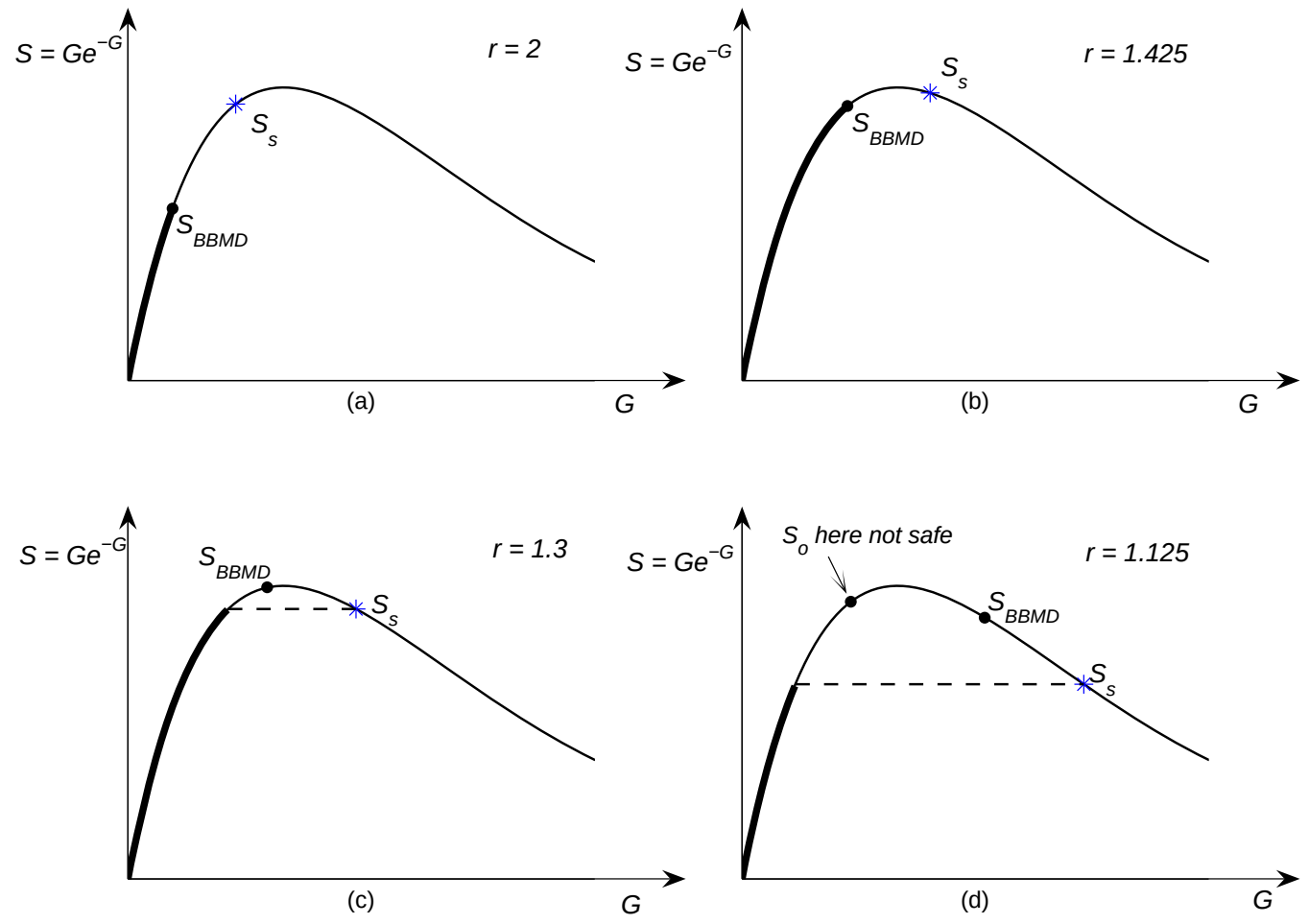

Fig. 3. Relative positions of $(G_{BBMD}, S_{BBMD})$ and $(G_s, S_s)$ on the $S = Ge^{-G}$ curve, and the associated feasible regions for bounded mean-delay, non-saturated operation (darkened lines), for (a) $r = 2$ (b) $r = 1.425$ (c) $r = 1.3$ (d) $r = 1.125$.

In Fig. 3(a), both $(G_{BBMD}, S_{BBMD})$ and $(G_s, S_s)$ are to the left of the peak of the *S*-*G* curve, with $S_{BBMD} < S_s$. According to Observation 2, the feasible region for bounded-mean-delay operation is to the left of $(G_{BBMD}, S_{BBMD})$, as shown in the figure. According to Observation 1, this region is also within the non-saturated operating region. Overall, bounded-mean delay and non-saturated operation can be ensured by limiting the offered load $S_o < S_{BBMD}$. For $S_o$ between $S_{BBMD}$ and $S_s$, the system is non-saturated but the mean delay is unbounded.

As *r* decreases, we have the situation in Fig. 3(b), where $(G_{BBMD}, S_{BBMD})$ is to the left and $(G_s, S_s)$ is to the right of the peak of the *S*-*G* curve, with $S_{BBMD} < S_s$. Again, with the same argument as for Fig. 3(a) above, non-saturation and bounded mean delay can be ensured by limiting $S_o < S_{BBMD}$. Also, for $S_o$ between $S_{BBMD}$ and $S_s$, the system is non-saturated but the mean delay is unbounded.

As $r$ decreases further, we have the situation in Fig. 3(c), where $(G_{BBMD}, S_{BBMD})$ is to the left and $(G_s, S_s)$ is to the right of the peak, but $S_{BBMD} > S_s$. Decreasing $r$ even further leads us to Fig. 3(d), where both $(G_{BBMD}, S_{BBMD})$ and $(G_s, S_s)$ are to the right of the peak, with $S_{BBMD} > S_s$. For both of these cases, limiting $S_o < S_s$ will ensure non-saturation and bounded mean delay.

For the two cases in Fig. 3(c) and Fig. 3(d), it may appear at first glance that we could load the system with $S_o > S_s$ and even $S_o > S_{BBMD}$ while ensuring bounded mean delay operation. To see this argument, suppose that we have an $S_o$ as shown in Fig. 3(d). According to the argument in the paragraph immediately below Observation 2, at this $(G_o, S_o)$, $p_c r^2 < 1$, satisfying the bounded mean-delay condition. In the following paragraph, we argue that it is in fact not "safe" to load the system with $S_o > S_s$.

When $S_o > S_s$, there is the danger of the system running into saturation, at which point $E[D]$ will go to infinity because the saturation throughput $S_s$ cannot keep up with the input rate $S_o$. That is, the equilibrium of the system as assumed in our local analysis in Section 3.3 does not apply any more. In a simulation experiment, for a situation such as that depicted in Fig. 3(d), we intentionally caused the system to go into saturation with a sudden increase in the offered load, and then decreased the offered load back to the $S_o$ shown in the figure. The simulation results show that $E[D]$ becomes unbounded thereafter. In other words, such an $S_o$ which is larger than $S_s$ is not a "safe" offered load, and it is obtained with an a priori assumption of equilibrium and non-saturation. If the system is already in saturation, $E[D]$ is unbounded for such an $S_o$ and cannot recover. On the other hand, in the simulation experiment, if we decreased the offered load further to below $S_s$, then the system did clear up and $E[D]$ became bounded. Indeed, what we observed was the "quantum jump" phenomenon discussed in Section 3.3.1 as $S_o$ crosses $S_s$. Thus, $S_o < S_s$ is safe.

Combining the descriptions of all four cases above, we thus arrive at Observation 4 below:

**Observation 4:** The feasible region for $(G_o, S_o)$ in terms of bounded mean delay and non-saturated operation is the intersection of the two feasible regions in Observations 1 and 2.

We define the "safe" bounded-mean-delay throughput as follows to correspond to Observation 4:

$$S_{SBMD}(r) = \min\left[S_{BBMD}(r),\ S_s(r)\right] = \min\left[\frac{r^2-1}{r^2}\ln\left(\frac{r^2}{r^2-1}\right),\ \frac{r-1}{r}\ln\left(\frac{r}{r-1}\right)\right] \quad (19)$$

**Finite-*N* Case : $S_{SBMD}$ as function of $r$, $r_0$, $N$**

We now consider the finite-$N$ case. The mechanic of the argument is similar to the $N \to \infty$ case. It can be shown that Observations 1, 2, and 4 remain intact on the finite-$N$ $S$-$G$ curve, $S = G\left(1 - G/N\right)^{N-1}$. However, Observation 3 may not be valid, as explained in the next paragraph. As a result, we cannot simply say $S_{SBMD}(r, r_0, N) = \min\left[S_{BBMD}(r, r_0, N),\ S_s(r, r_0, N)\right]$. Nevertheless, Observation 4 can still be used to identify the feasible region for $S_o$.

For finite $N$, the three equations $S_o = G_o\left(1 - G_o/N\right)^{N-1}$, $p_c r^2 = 1$, and $p_c = 1 - S_o/G_o$ yield $S_{BBMD} = N(1 - 1/r^2)\left[1 - \left(1 - 1/r^2\right)^{1/(N-1)}\right]$. Meanwhile, $S_s(r, r_0, N)$ has no closed form but can be found numerically from (5). Numerically, we find that as $r$ decreases, it is possible for $(G_{BBMD}, S_{BBMD})$ to overtake $(G_s, S_s)$ so that it moves to the right of $(G_s, S_s)$. With respect to the situation in Fig. 3(d), if $(G_{BBMD}, S_{BBMD})$ is to the right of $(G_s, S_s)$, it is also below $(G_s, S_s)$. As a result, the intersected feasible region mentioned in Observation 4 includes the region where $S_{BBMD} \le S_o < S_s$, in addition to the region where $S_o < S_{BBMD}$. In this case, $S_{SBMD}(r, r_0, N) = S_s(r, r_0, N)$ rather than $S_{SBMD}(r, r_0, N) = \min\left[S_{BBMD}(r, r_0, N),\ S_s(r, r_0, N)\right]$.

## 4. Effects of Backoff Factor *r*

The analysis in the preceding section hinted that the backoff factor $r$ may have a significant impact on the system performance. This section is devoted to a detailed study of the effect of $r$.

### 4.1 Maximum SBMD Throughput

Let us now examine how $S_{SBMD}$ varies as $r$ is varied. We focus on the asymptotic $N \to \infty$ case here. Similar argument applies to the non-asymptotic case although the equations are more complicated. Fig. 4 plots $S_{SBMD}(r)$, $S_{BBMD}(r)$, and $S_s(r)$ versus $r$ according to (17), (18), and (19).

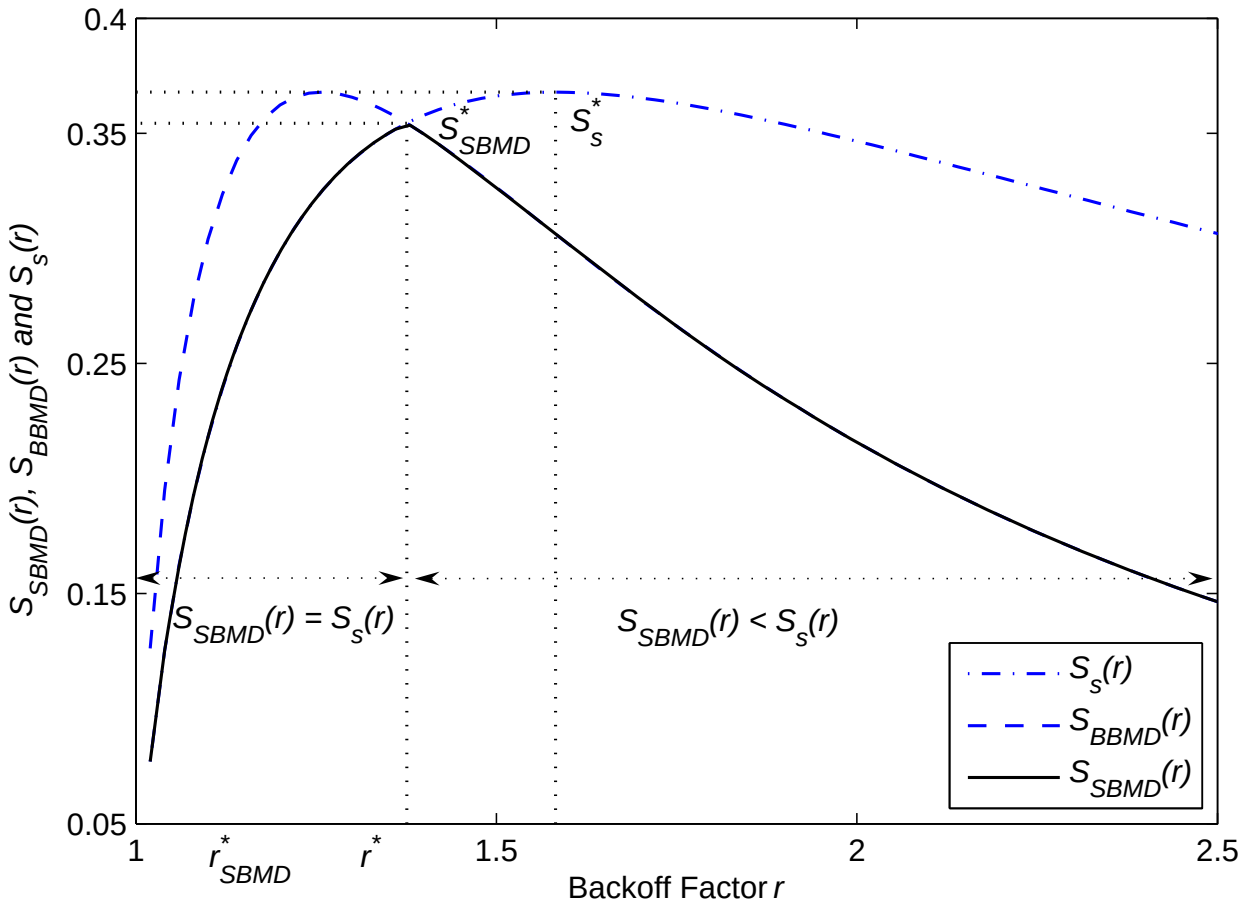

Fig. 4. $S_{SBMD}(r)$, $S_{BBMD}(r)$, and $S_s(r)$ versus $r$ for the case of $N \to \infty$.

For $r > 1.3757$, $S_{BBMD}(r) < S_s(r)$; and for $r \le 1.3757$, $S_{BBMD}(r) \ge S_s(r)$. Specifically, the $r$ which maximizes $S_{SBMD}(r)$ is $r^*_{SBMD} = 1.3757$, which is obtained by setting $S_{BBMD}(r) = S_s(r)$:

$$\frac{{r^*_{SBMD}}^2 - 1}{{r^*_{SBMD}}^2} \ln\left(\frac{{r^*_{SBMD}}^2}{{r^*_{SBMD}}^2 - 1}\right) = \frac{r^*_{SBMD} - 1}{r^*_{SBMD}} \ln\left(\frac{r^*_{SBMD}}{r^*_{SBMD} - 1}\right) \qquad (20)$$

Note that $r^*_{SBMD} \ne r^*_s = e/(e-1)$, where the $r^*_s$ is the value of $r$ value that maximizes the saturation throughput $S_s(r)$. The maximum saturation throughput $S^*_s = S_s(r^*_s) = e^{-1} = 0.3679$. However, $S_{SBMD}(r^*_s) = 0.3063$, which is 17% below $S^*_s$. That is, if we set $r = r^*_s$, the offered $S_o$ load must be at least 17% below the saturation throughput $S^*_s$ to ensure bounded delay operation.

The binary backoff factor of $r = 2$ is assumed in the majority of prior work, and in many practical multiple-access networks such as the Ethernet and WiFi. For slotted Aloha, the corresponding saturation throughput $S_s(2) = 0.3466$ is reasonably close to $S^*_s = 0.3679$, and one could hardly raise objection to adopting $r = 2$ on the basis of saturation throughput. However, if bounded mean delay is desired, we have $S_{SBMD}(2) = 0.2158$. That is, there is a drastic 41% penalty with respect to $S^*_s$. Therefore, $r = 2$ is a bad choice from the delay consideration.

Fortunately, the maximum SBMD throughput, obtained by setting $r = r^*_{SBMD} = 1.3757$, is rather close to $S^*_s$. Specifically, $S^*_{SBMD} = S_{SBMD}(r^*_{SBMD}) =$ 0.3545. The penalty with respect to $S^*_s$ is only less than 4%. Overall, we conclude that using the proper $r$ is important to ensuring a good throughput under the bounded-delay requirement, perhaps more so than when saturation throughput is the only concern. This can be seen from Fig. 4, which shows that $S_{SBMD}(r)$ rises and falls much more sharply with $r$ than $S_s(r)$ does.

## 4.2 Mean Delay versus Offered Load

Fig. 5 plots $E[D]$ versus $S_o$ for the case of $N = 30$. Numerically, $E[D]$ is obtained as follows. For a given $S_o$, we compute $G_o$ from $S_o = G_o(1 - G_o/N)^{N-1}$. Recall from the discussion in Section 3.3.1 that this will yield two solutions, $G_{o,l}$ and $G_{o,r}$ but that the smaller $G_{o,l}$ is the correct operating point. We substitute $p_c = (G_{o,l} - S_o)/G_{o,l}$ and $\lambda_o = S_o/N$ into (13) to find $E[D]$.

In Fig. 5(a), $(r_0, r, N) = (10,\ 1.582,\ 30)$. For this case $S_{BBMD} = 0.3140 \ < 0.3675 = S_s$. This case corresponds to the situation in Fig. 3(a). $S_{SBMD}$ is limited by $S_{BBMD}$ rather than the saturation throughput $S_s$. The solid line in Fig. 5(a) is the result of from numerical analysis. The cross points are simulation results of the proxy system in which the dynamic of a single node is simulated with fixed $p_c = (G_{o,l} - S_o)/G_{o,l}$ computed numerically. The dotted points are simulation results of the real system. The results are consistent in that for offered load $S_o$ near $S_{BBMD}$, $E[D]$ begins to build up quickly.

An interesting observation is that near $S_{BBMD}$, the simulated $E[D]$ does not converge in either the proxy or the real system. Noteworthy is the fact that the simulation results of the proxy system can fluctuate below and above the numerical results of the proxy system, although the simulation experiment simulates exactly the same proxy system as that in the analysis. In other words, this non-convergence is not due to the proxy system not being able to approximate the real system well. On the contrary, the proxy system suggests that similar non-convergence may happen in the real system, which is borne out by our simulation results. In fact, for the same $S_o$ near $S_{BBMD}$, different simulation runs will produce rather different $E[D]$ even if we let each run lasts a long time. The underlying cause of such non-convergence will be further discussed in Section 5.2.

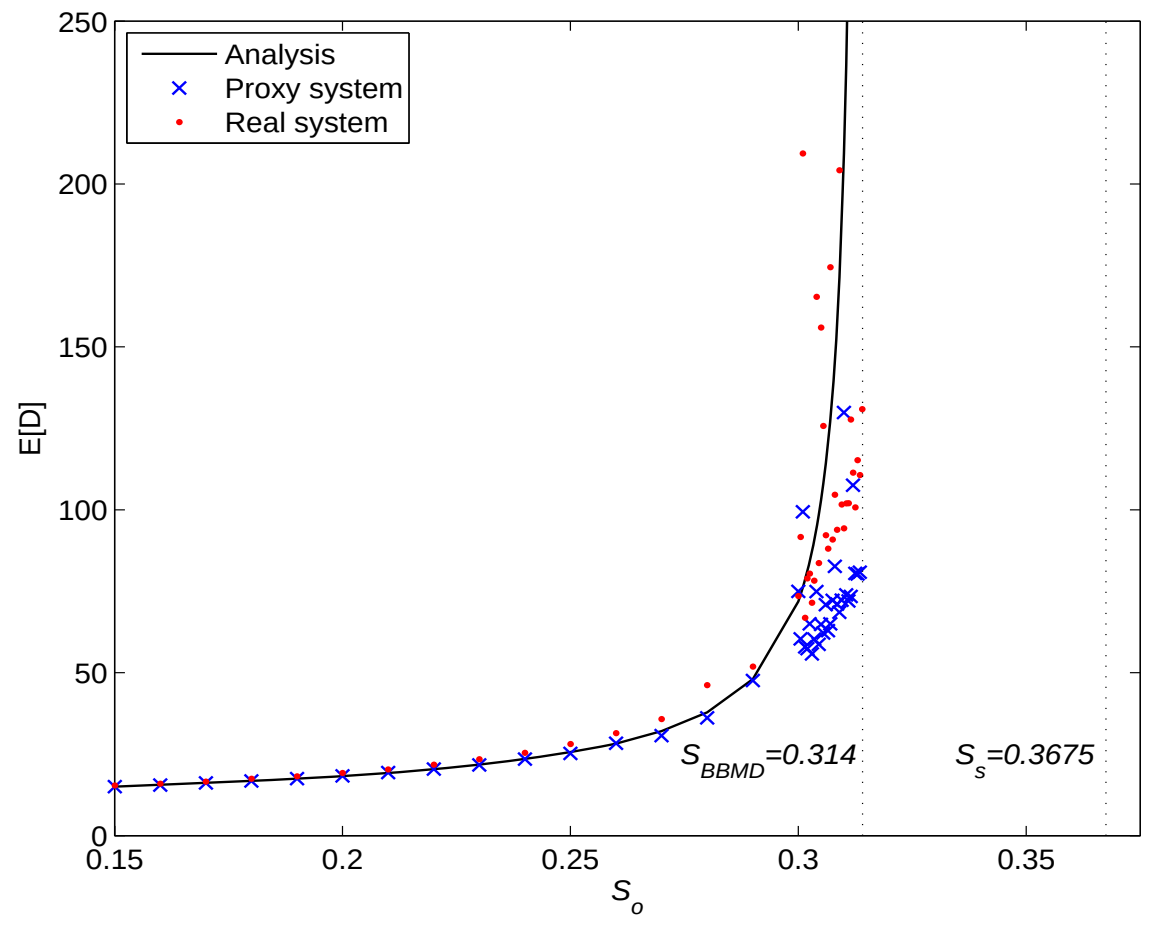

(a)

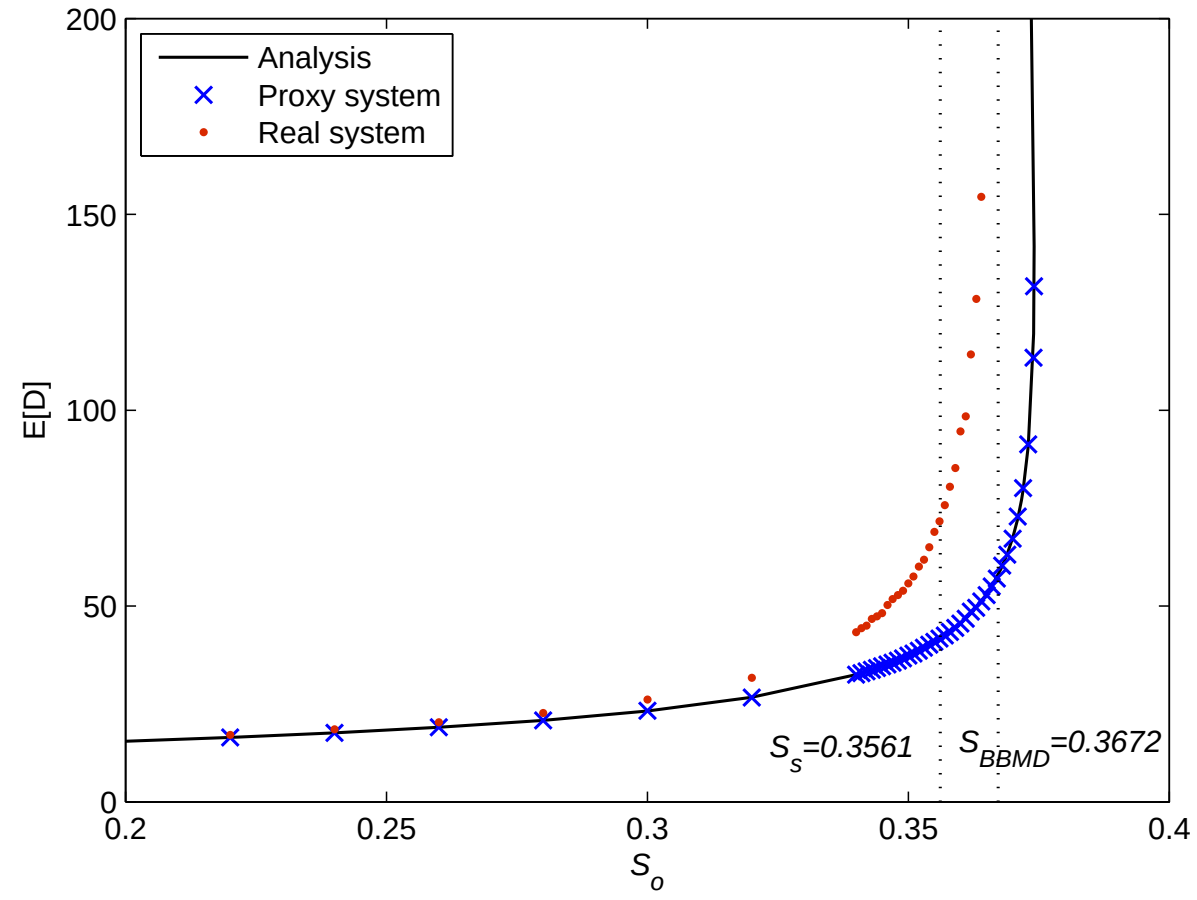

(b)

Fig. 5. $E[D]$ versus $S_o$ for (a) $(r_0, r, N) = (10,\ 1.582,\ 30)$; (b) $(r_0, r, N) = (10,\ 1.200,\ 30)$.

We have also simulated the setting $(r_0, r, N) = (10,\ 2,\ 30)$, an interesting case because $r = 2$ was assumed in many prior studies. This case also corresponds to the situation in Fig. 3(a). We omit the plot of the curve here to conserve space. The quality results are similar to those described in the two paragraphs above for the $(r_0, r, N) = (10,\ 1.582,\ 30)$ setting. This time, however, $E[D]$ goes to infinity earlier at $S_o = S_{BBMD} = 0.2221$, as predicted analytically. In other words, $r = 2$ is not a good setting from the delay perspective because the offered load will be much limited.

In Fig. 5(b), $(r_0, r, N) = (10,\ 1.200,\ 30)$ . For this case $S_{BBMD} = 0.3762 > 0.3561 = S_s$ , and $S_{SBMD}$ is limited by $S_s$ rather than $S_{BBMD}$ . This is a rather subtle case corresponding to Fig. 3(d), in which both $S_{BBMD}$ and $S_s$ are to the right of the peak of the *S*-*G* curve. It is possible to load the system with $S_o$ above $S_{BBMD}$ and yet satisfy the convergence condition as dictated by (14). The analytical and simulation results in Fig. 5(b) confirm that. Such an $S_o > S_{BBMD}$ that has a finite $E[D]$ in Fig. 5(b), however, may be "unstable" in another sense. In a simulation experiment, we used an even larger $S_o$ to jolt the system into saturation, and then decreased $S_o$ back to the original value $S_o > S_{BBMD}$ . The system did not get out of saturation and $E[D]$ became unbounded and (14) is not satisfied thereafter. That is, the bounded $E[D]$ as in Fig. 5(b) would elude us once the system is saturated. Even if we did not jolt the system into saturation as above, the system may eventually evolve to the saturation state with a constant $S_o > S_{BBMD}$ . How soon it does that depends on how close $S_o$ is to the peak of the *S*-*G* curve. The intricate dynamic on how long the system can remain stable at an offered load above $S_{BBMD}$ is an interesting subject for further research work.

## 5. Starvation

Starvation occurs when some nodes do not get to transmit their packets for an excessively long time. This may happen, for example, when the nodes back off exponentially to a large backoff stage. Other nodes with a smaller backoff stage will then hog the channel. As far as we know, the "qualitative" observation of the starvation phenomenon was first made in [2] (although under a different backoff protocol). The authors attributed the discrepancy between their simulation and analytical results to starvation. Left open are three major outstanding issues:

(i) What is the appropriate "quantitative" definition of starvation? To study starvation systematically, we need a starvation metric that is measurable, much like delay is measurable.

(ii) Why does starvation lead to a discrepancy between simulation and analytical results? What is the root cause of this phenomenon?

(iii) How are system parameters $r$, $r_0$, $N$, $S_o$ in our system model related to starvation quantitatively?

Sections 5.1, 5.2, and 5.3 address (i), (ii), and (iii), respectively.

### 5.1 Definition of Starvation

Fundamentally, starvation is related to HOL service. There is a vague notion that when a HOL packet does not receive service for a long time, the associated queue is then starved. Thus, an attempt to define starvation quantitatively could focus on the property of the HOL service time $X$ .

Consider all the busy times of all nodes. Suppose that we randomly choose a node and a point within its busy times to observe the service time of the HOL packet into which the random point falls. Then the random variable that we observe is not *X*. It is another random variable *Y*, whose probability distribution $P_Y(y) = \Pr[Y = y]$ is related to the probability distribution of $X$, $P_X(x) = \Pr[X = x]$, by $P_Y(y) = yP_X(y)/E[X]$. The weight $y$ is due to the fact that the random point we sample is proportionately more likely to fall within a long service time than a short service time; and the denominator $E[X]$ is a normalization factor so that $P_Y(y)$ sums up to one. This "node-centric" sampling makes sense as far as starvation is concerned, since we are interested in whether a busy node is suffering from a long service time at a randomly chosen time.

A number of definitions of starvation around *Y* are possible. For example, we could say that there is no starvation if and only if $\Pr[Y > y_{\text{target}}] < \varepsilon$ for some $y_{\text{target}} > 0$ and $\varepsilon > 0$ ; another possibility is $E[Y] < \bar{Y}_{\text{target}}$ for some target mean $\bar{Y}_{\text{target}} > 0$ . For the rest of this paper, we adopt the simple definition that requires $E[Y]$ to be finite:

**Definition of Non-Starvation:** A system is non-starved if and only if $E[Y]$ (hence $E[X^2]$) is finite.

That $E[Y]$ is finite does not mean it is small. The implicit understanding behind this definition is that whatever condition we come up with that can meet the finite $E[Y]$ requirement, we need to use a condition that is somewhat tighter in actual implementation. This is analogous to the definition of $S_{SBMD}$ , where we need to make $S_o$ smaller than $S_{SBMD}$ by a sufficient amount if we want to meet certain targeted mean delay (i.e., we cannot simply set $S_o = S_{SBMD}$ ).

With this definition, we can now relate the condition for non-starvation to the condition for bounded mean delay in a non-saturated system. Mathematically, it can be easily shown from $P_Y(y) = yP_X(y)/E[X]$ that $E[Y]$ is bounded if and only if $E[X^2]$ is bounded. According to (13), if $E[X^2] = X''(1) + X'(1)$ is not bounded, then $E[D]$ is also not bounded. The practical significance and interpretation is as follows. When $E[X^2]$ is large, not only will the delay performance be bad, the performance among different nodes may also vary widely because some are starved while others are not.

Our definition of starvation allows us to unite the notions of non-starved operation and bounded-mean-delay operation, since a root cause giving rise to both of them is the same: large $E[X^2]$ .

### 5.2 Starvation and Non-Convergence of Simulations

This section explores why non-convergence of simulation results happens to occur whenever the system is starved, a phenomenon observed in [2] as well as in our simulation experiments.

Underlying this phenomenon is a fundamental cause: the immeasurability of performance when starvation occurs, as explained below.

**Saturated Case**

Starvation can occur in a saturated or non-saturated system. We first focus on the saturated case. Suppose we want to measure the average service time $E[X]$ at saturation (note: $S_s = 1/E[X]$ by Little's law). In the following, we argue that for a starved system, $E[X]$ cannot be estimated accurately. For our measurement, imagine that we perform $m$ experiments, $m \gg 1$. Each experiment $j \in \{1, 2, ..., m\}$ is conducted over a long time so that we could gather the HOL service times of $n_p \gg 1$ packets of a particular queue. For each trace *j*, we can compute the average service time as

$$\bar{X}_j = \frac{\sum_{i=1}^{n_p} X_{j,i}}{n_p}, \quad j = 1, ..., m \tag{21}$$

where $X_{j,i}$ is sample $i$ of trace *j*. From the large set of $m$ experiments, we have $m$ samples of $\bar{X}_j$ from (21). From the samples, we can then construct the probability density of $\bar{X}_j$, $f_{\bar{X}_j}(\bar{x})$. Let us make $n_p$ very large for each of the experiments. We wish that the Law of Large Numbers would then apply, and the spread of this density would then become very narrow. If so, we could estimate $E[X]$ accurately by defining $E[X] = \bar{X}_j$ for any $j$ since $\bar{X}_j$ for different $j$ converges; if not, we really do not know which $\bar{X}_j$ is to be believed, and a definitive measure of $E[X]$ would elude us. Note the caveat that if $\bar{X}_j$ does not converge as $n_p$ increases, $E[X]$ alternatively defined as $E[X] = (\sum_j \bar{X}_j)/m$ does not converge either, since this is equivalent to increasing the sample size $n_p$, which does not help.

We show in the following that if the system is starved and $E[X^2]$ is unbounded, then $E[\bar{X}_j^2]$ is unbounded; hence, $f_{\bar{X}_j}(\bar{x})$ does not "narrow" with large $n_p$. The expectation in (22) below is the ensemble average over a large number of experiments.

$$E[\bar{X}_j^2] = \frac{1}{n_p^2} E[(\sum_{i=1}^{n_p} X_{j,i})^2] \geq \frac{1}{n_p^2} E[\sum_{i=1}^{n_p} X_{j,i}^2] = \frac{E[X^2]}{n_p} \tag{22}$$

Thus, $E[\bar{X}_j^2]$ is unbounded if $E[X^2]$ is unbounded. Of course, in experiments, our measurement is time-limited by the duration of our experiment, and we will not observe $\bar{X}_j^2$ to be infinite. Nevertheless, the above points out that it is likely that $\bar{X}_j$ will not converge in experiments.

Fig. 6 presents our experimental results. We set $(r_0, r, N) = (10,\ 1.582,\ 15)$, a starved case where $E[X^2]$ is unbounded. Fig. 6 (a) and (b) are the results of the real system and proxy system, respectively. The number of experiments in each set is $m = 5$. For each trace, $\bar{X}_j$ of one queue is measured as a function of $n_p$, as per (21). Specifically, as each packet departs from the queue, $n_p$ increases by one, and $\bar{X}_j$ is recomputed to take the statistic of this packet into account. For the experiments of the proxy system, we first compute the $p_c$ as a function of $r_0$, $r$, $N$ from (4), and then use this $p_c$ to simulate the Markov chain associated with a queue.

In both the real and proxy systems, there is a spread of $\bar{X}_j$ across the $m$ experiments, and that they do not converge to a common value as $n_p$ increases. In contrast, for the case of $(r_0, r, N) = (10,\ 1.2,\ 15)$, a non-starved case, $\bar{X}_j$ converges to a common value as $n_p$ increases (the results of this set of experiments are not shown here to conserve space). A point worth emphasizing is that such non-convergence is not related to the proxy system not accurately approximating the dynamic in the real system, since non-convergence occurs in both systems.

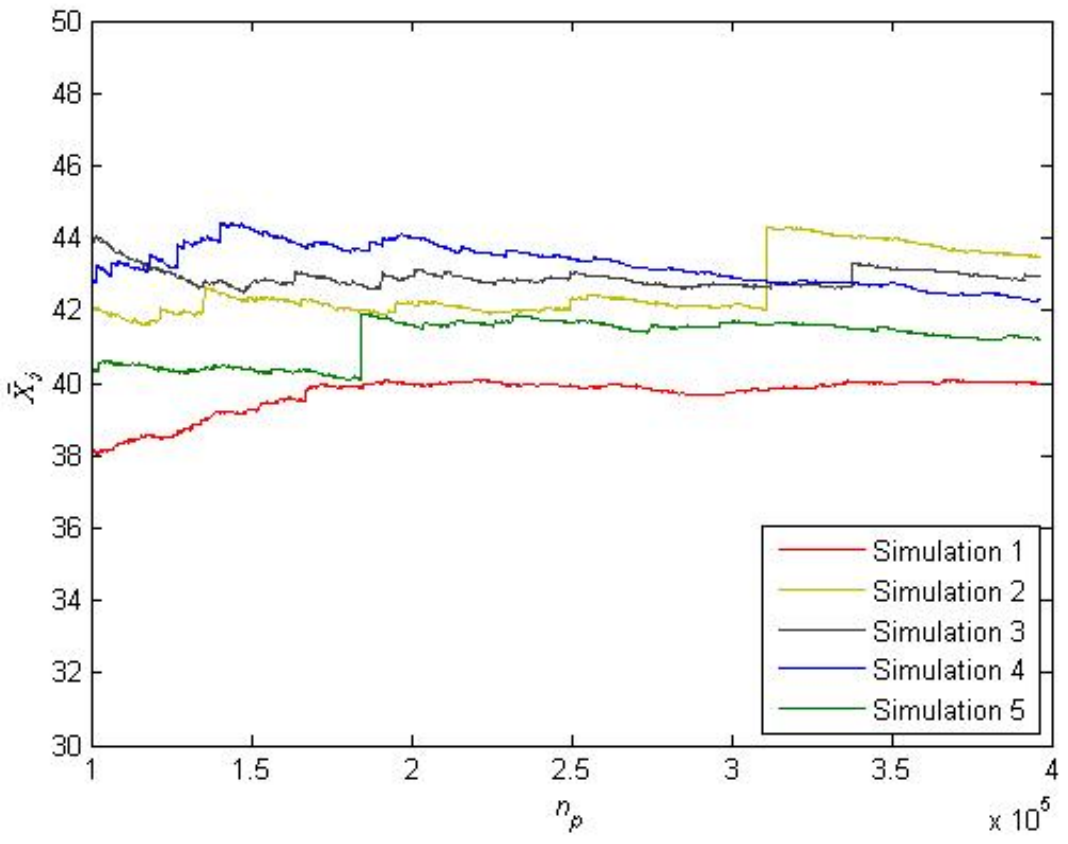

(a)

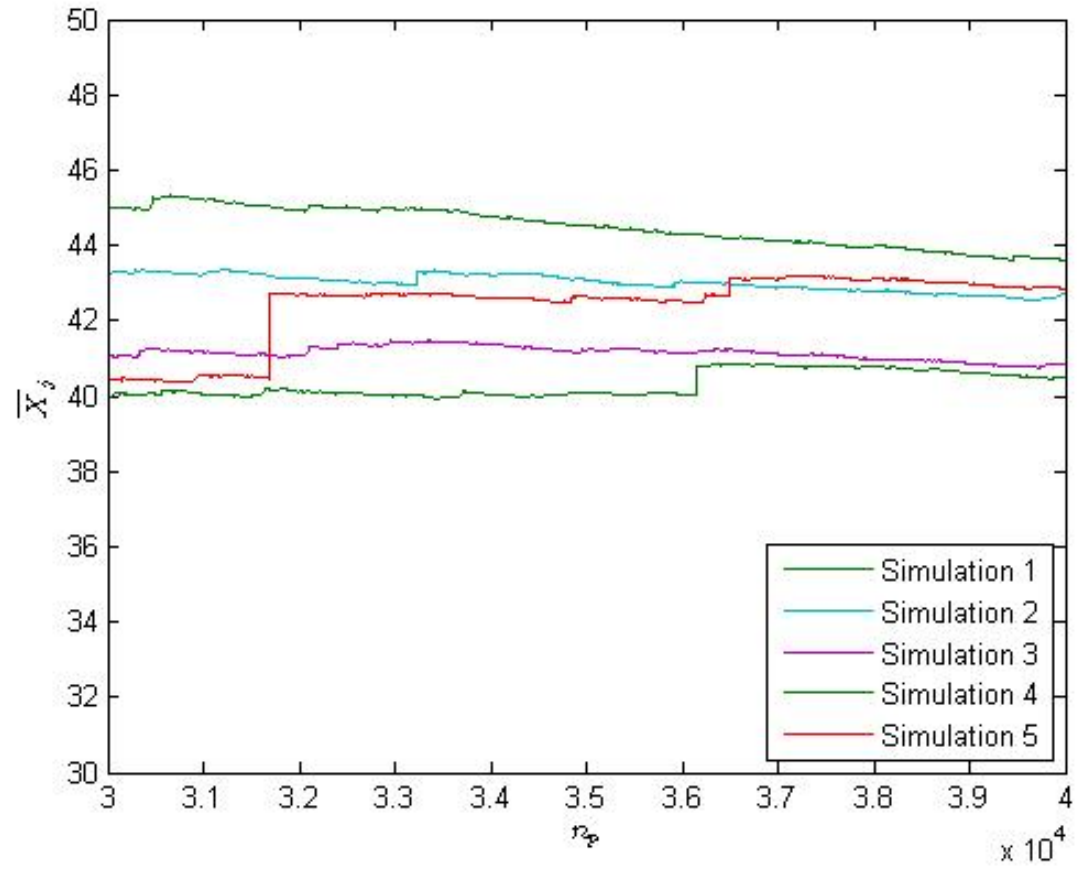

(b)

Fig. 6. Measurement of $\bar{X}_j$ as number of samples $n_p$ increases for parameter setting, $(r_0, r, N) = (10,\ 1.582,\ 15)$ in (a) real system; (b) proxy system.

Since $\bar{X}_j$ does not converge, neither does the average throughput of the queue (if we measure average throughput of a queue as $n_p \big/ \sum_{i=1}^{n_p} X_{j,i} = 1 \big/ \bar{X}_j$ ). Indeed, in our experiments, we observe the throughputs of different queues are quite different even if we average the throughputs over a long stretch of time. Unfairness tends to persist.

In summary, the phenomena of starvation and non-convergence of measured performance results are intricately tied, and they have the same root cause: unbounded $E[X^2]$.

**Non-saturated Case**

The above has focused on the saturated case. Non-convergence also occurs in the non-saturated case. In the non-saturated case, the offered load $S_o$ is a factor as to whether starvation occurs.

Besides the non-convergence of measured $E[X]$, which occurs when $E[X^2]$ is unbounded, the measured $E[D]$ may not converge either. For the same reason that $E[X^2]=\infty$ does not allow converged measurement of $E[X]$, $E[D^2]=\infty$ does not allow converged measurement of $E[D]$ either. It can be shown from $D^*(s)$ in (8) that $E[D^2]$ goes to infinity before $E[X^2]$ does (omitted here to conserve space). This is borne out by Fig. 5(a) in which the measured $E[D]$ begins to diverge before $S_o$ reaches $S_{BBMD}$. Again, the non-convergence of the measured $E[D]$ has nothing to do with the inaccuracy of the proxy system with respect to the real system. Even for the proxy-system simulation, as shown in Fig. 5(a), there is a spread in the measured $E[D]$ due to the fundamental reason of immeasurability.

## 5.3 Impact of System Parameters on Starvation

We now investigate how system parameters affect starvation.

**Saturated Case**

For the study of the saturated case, we note that the expression of $X(z)$ in (12) for the non-saturated case is also valid for the saturated case because it is parameterized on $p_c$. Following (12), (14) indicates that bounded $E[X^2]$ requires $p_c r^2 < 1$. We just need to be careful to substitute the $p_c$ obtained from the global analysis of the saturated case rather than that from the non-saturated analysis.

For fixed $r$, $r_0$, it turns out that starvation sets in when the number of nodes $N$ is beyond a certain value. Here, we are interested in this critical value of $N$. It can be shown from (4) that $N$ is an increasing function of $p_c$ for $r>1$. Rearranging (4), we have

$$N = 1 + \frac{\ln(1-p_c)}{\ln\left(1-\dfrac{1-p_c r}{r_0(1-p_c)}\right)} \qquad (23)$$

Substituting $p_c < 1/r^2$ (condition for bounded $E[X^2]$) gives

$$N < 1 + \frac{\ln\left(1-1/r^2\right)}{\ln\left(1-\dfrac{1-1/r}{r_0(1-1/r^2)}\right)} = \frac{\ln\left(\dfrac{r}{r-1}\right)-\ln\left(1+\dfrac{1}{r}-\dfrac{1}{r_0}\right)}{\ln\left(\dfrac{r+1}{r}\right)-\ln\left(1+\dfrac{1}{r}-\dfrac{1}{r_0}\right)} \triangleq N_s^* \qquad (24)$$

where $N_s^*$ is the critical value we seek. Note that $N_s^*$ increases with $r_0$ but decreases with $r$.

To illustrate the phenomenon of starvation, we present in Fig. 7 a simulation trace of a real system with $(r_0, r, N) = (10,\ 1.2,\ 30)$. According to (24), this parameter setting will result in starvation. We simulated a total of 20 million time slots, and examined one particular node. Specifically, we looked at the number of cleared packets of the node within each time window of $7{,}500 \approx 100N/S_s$ slots. Thus, the expected number of cleared packets per time window is 100. Fig. 7 plots the number of cleared packets for successive time windows. Note that besides the large spread in the number of cleared packets, there are two occasions during which the node receives no service at all for a very long time. The first occasion lasts for 1.1 million slots, and the second occasion lasts for 0.33 million slots.

Before concluding the discussion here, we would like to point out that the study of the saturated case is particularly relevant to the scenario in which each node is a TCP source. TCP is a greedy transport-layer protocol. For long-lasting TCP applications, such as FTP or P2P File Sharing, a TCP connection will attempt to keep the queue at the MAC layer occupied at all times, thus causing the system to operate in saturation. Relationship (24) allows us to determine the maximum number of active nodes in an Aloha network before starvation sets in, and how this number depends on $r$ and $r_0$. When the number of active nodes is too large, some of them will be starved, leading to unfairness. Generally, smaller $r$ is more robust against starvation (see Fig. 8). However, bear in mind that the overall saturation throughput will also go down if $r$ is too small (according to (5)) Thus, there is a tradeoff between system throughput and fairness. Relationships (5) and (23) allow us to engineer the right balance by tuning $r$ and $r_0$.

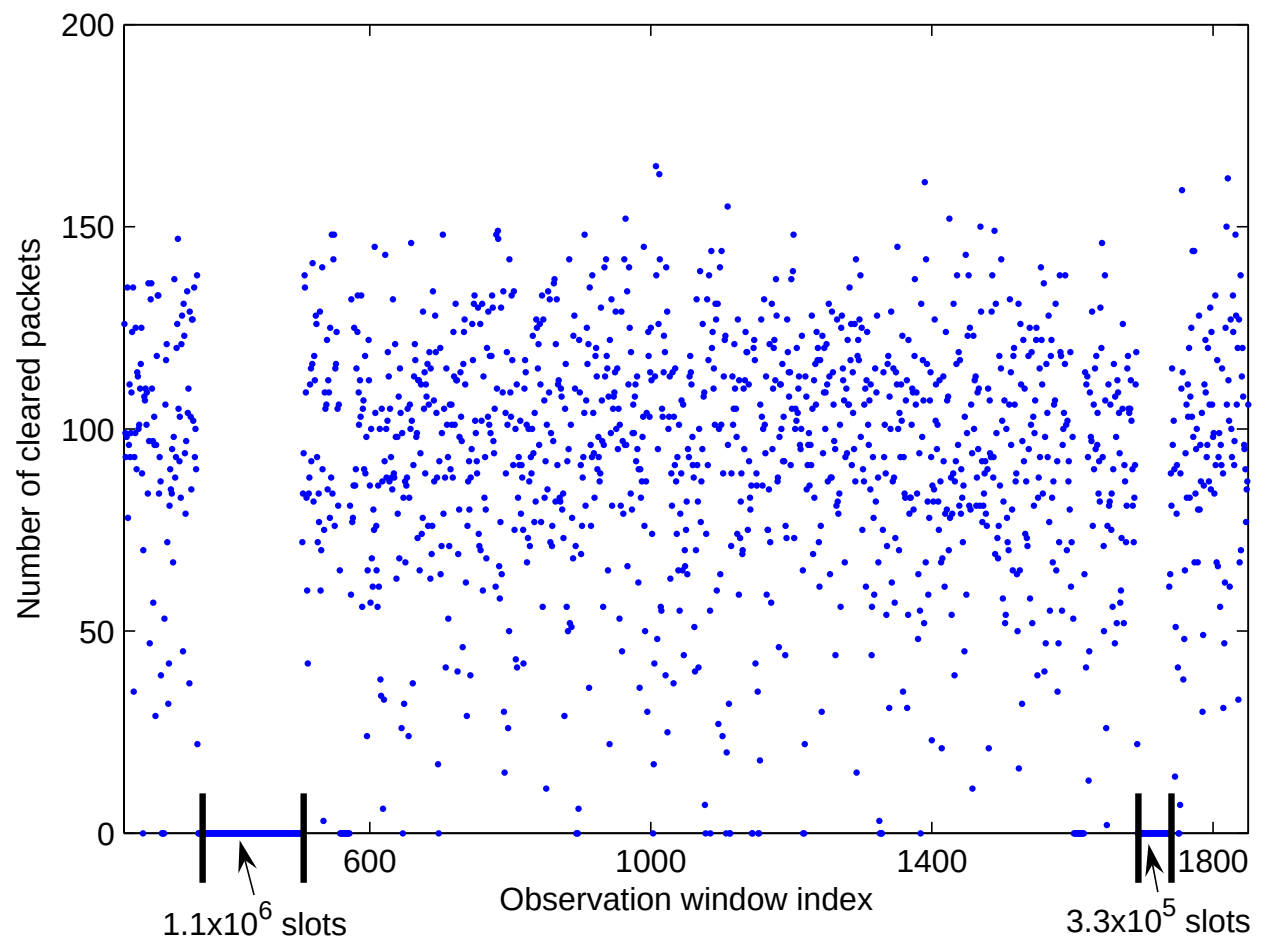

Fig. 7. Illustration of Starvation: Number of cleared packets of a node in successive time windows for a real system with $(r_0, r, N) = (10,\ 1.2,\ 30)$. Each time window consists of 7,500 time slots.

**Non-saturated Case**

For the non-saturated case, the offered load $S_o$ is a design parameter in addition to $r,\ r_0,\ N$ . In general, there is a feasible region for non-starved operation within the space $(r,\ r_0,\ N,\ S_o)$. Unlike in the saturated case, in the non-saturated case the additional degree of freedom in $S_o$ allows us to support large $N$. For any $N$, we could make $S_o$ small enough to avoid starvation.

Consider the asymptotic $N \to \infty$ case, and suppose that we load the system with $S_o < S_s$ to ensure non-saturated operation. The feasible region is then governed by $S_o < S_{SBMD}(r)$ in (19), which is independent of $r_0$ as well as $N$. Essentially, the feasible region for non-starved operation is the same as that for bounded-mean delay operation. This observation again ties together the notions of bounded-mean delay and non-starvation. The largest possible offered load for non-starved operation is therefore $S_{SMBD}(r) = 0.3545$, obtained when $r = r^*_{SBMD} = 1.3757$ (see Section 4.1).

## 6. Conclusions

We have presented an analytical framework for the study of queuing delay and starvation in the slotted Aloha network operated with the exponential backoff protocol. Based on the framework, we have derived the dependency of queuing delay and non-starved operation on the system parameters, including the backoff factor, the initial transmission probability, and the number of nodes in the network.

With respect to delay performance, we showed that the system offered load $S_o$ must be below a "safe-bounded-mean-delay throughput", $S_{SBMD}$, in order that the mean delay is bounded. Specifically, for the case in which the number of nodes is large, the sustainable offered load must be limited as follows:

$$S_o < S_{SBMD} \triangleq \min\left[\frac{r^2-1}{r^2}\ln\left(\frac{r^2}{r^2-1}\right),\ \frac{r-1}{r}\ln\left(\frac{r}{r-1}\right)\right] \qquad (25)$$

where $r$ is the backoff factor. The first term in the min[ ] function is due to the need to bound the service-time variance, and the second term is the saturation throughput. Worth noting from (25) is that $S_{SBMD}$ is smaller than or equal to the well-known saturation throughput $S_s$. This means that we cannot automatically assume we could load the system with offered load up to the saturation throughput when delay performance is a concern.

With respect to starvation, for a non-saturated system, we argued that the conditions for bounded mean delay and non-starvation are one of the same, thus uniting these two notions. For a large Aloha network, for example, limiting the offered load to below the $S_{SBMD}$ given in (25) can ensure bounded-mean-delay and non-starved operations.

Starvation is also a concern in a saturated system. Saturation can occur, for example, when the applications at the nodes run the TCP transport protocol on top of the MAC protocol. TCP connections, being greedy in nature, will keep the queues occupied at all time, thus saturating the system. Unlike in the non-saturated case, in the saturated case the number of nodes $N$ rather that the offered load $S_o$ must be limited. The bound on $N$ is given by $N_s^*$ below:

$$N < N_s^* \triangleq \frac{\ln\left(\frac{r}{r-1}\right) - \ln\left(1+\frac{1}{r}-\frac{1}{r_0}\right)}{\ln\left(\frac{r+1}{r}\right) - \ln\left(1+\frac{1}{r}-\frac{1}{r_0}\right)} \qquad (26)$$

where $1/r_0$ is the initial transmission probability.

A general conclusion is that delay and non-starved performance can be very sensitive to the system parameters; indeed, much more so than the saturation throughput is. Careful tuning of the system parameters is important. For example, consider a large Aloha. The maximum throughput is well known to be $e^{-1} = 0.3679$. The binary backoff factor $r = 2$ is assumed in many prior investigations and the corresponding saturation throughput is $S_s(2) = 0.3466$, which is close to the maximum of 0.3679. However, if we want to bound mean delay and prevent starvation, according to (25), the offered load $S_o$ must be below $S_{SBMD}(2) = 0.2158$, a drastic 41% lower than 0.3679. Therefore, setting $r = 2$ is not desirable from the standpoint of good delay and non-starvation performance, although it may achieve good saturation throughput. By tuning $r$ to $r^*_{SBMD} = 1.3757$, $S_{SBMD}$ can be maximized. The corresponding result is $S_{SBMD}(1.3757) = 0.3545$, which is less than 4% below 0.3679. Thus, $r = r^*_{SBMD}$ allows us to achieve good overall system throughput, good delay performance and non-starvation at the same time.

Last but not least, although a main focus of this paper is on mean delay, the analytical framework is general enough that higher moments of delay can also be studied using similar procedures propounded in this paper. Specifically, the Laplace Transform of delay in (8) can be used to generate higher moments of delay, and the three-step global-local-coupling analysis expounded in this paper can then be used to derive conditions needed to bound the higher moments.

Finally, two natural generalizations of the methods and results in this paper here are for carrier-sense multiple-access (CSMA) networks and networks with multiple-packet-reception (MPR) capability [13]. A companion paper of ours [14] is an attempt in that direction.

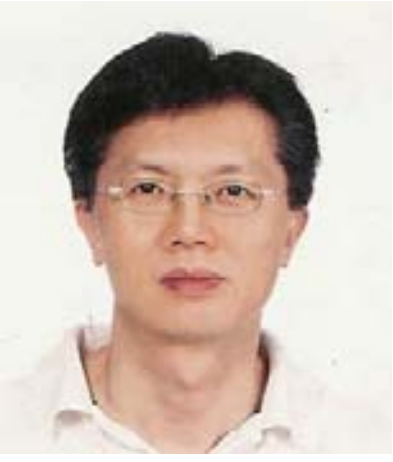

**Soung Chang Liew** (S'87–M'88–SM'92) received his S.B., S.M., E.E., and Ph.D. degrees from the Massachusetts Institute of Technology. From 1984 to 1988, he was at the MIT Laboratory for Information and Decision Systems, where he investigated Fiber-Optic Communications Networks. From March 1988 to July 1993, Soung was at Bellcore (now Telcordia), New Jersey, where he engaged in Broadband Network Research. He is currently Professor and Chairman of the Department of Information Engineering, the Chinese University of Hong Kong. Soung's current research interests include wireless networks, Internet protocols, and multimedia communications. Soung and his student won the best paper awards in *IEEE MASS 2004* and *IEEE WLN 2004*. Separately, TCP Veno, a version of TCP to improve its performance over wireless networks proposed by Soung and his student, has been incorporated into a recent release of Linux OS. Besides academic activities, Soung is also active in the industry. He co-founded two technology start-ups in Internet Software and has been serving as consultant to many companies and industrial organizations. Soung is Fellow of IEE and HKIE. Publications of Soung can be found in www.ie.cuhk.edu.hk/soung.

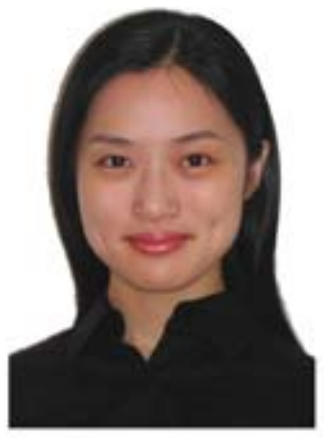

**Ying Jun (Angela) Zhang** (S'01-M'05) received BEng degree with Honors in Electronic Engineering from Fudan University, Shanghai China, in 2000, and Ph.D. degree in Electrical and Electronic Engineering from The Hong Kong University of Science and Technology in 2004. Since Jan. 2005, she has been with the Department of Information Engineering, where she is currently an assistant professor.
Dr. Zhang is on the Editorial Boards of IEEE Transactions on Wireless Communications and Wiley Security and Communications Journal. She has served as a TPC Co-Chair of Communication Theory Symposium of IEEE ICC 2009, Track Chair of ICCCN 2007, and Publicity Chair of IEEE MASS 2007. Her research interests include wireless communications and mobile networks, adaptive resource allocation, cross-layer design and optimization, wireless LAN, and MIMO signal processing.
Dr. Zhang won the Hong Kong Young Scientist Award 2006 as the only winner in the category of Engineering Science.

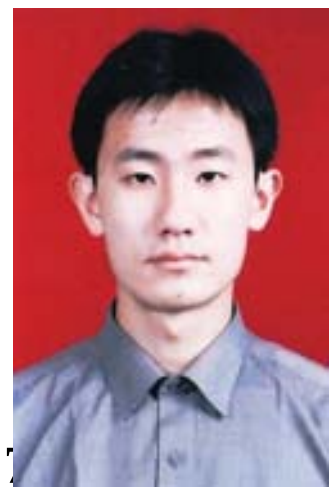

**Da Rui Chen** (S'07) received his B.Eng. degree in Information Engineering from Xi'an Jiaotong University, Xi'an, China, in 2005, and the M.Phil. degree in Information Engineering from The Chinese University of Hong Kong, Hong Kong, China, in 2007. Currently he is a research assistant in the Department of Information Engineering, The Chinese University of Hong Kong. His research interests include mobile and ad hoc networks, cross-layer design, and wireless MAC.